# Exploitation Business:
# Leveraging Information Asymmetry

Kwangseob Ahn
haebom@korea.ac.kr

## Abstract

This paper investigates the "Exploitation Business" model, which capitalizes on information asymmetry to exploit vulnerable populations. It focuses on businesses targeting non-experts or fraudsters who capitalize on information asymmetry to sell their products or services to desperate individuals. This phenomenon is also described as "profit-making activities based on informational exploitation," which thrives on individuals' limited access to information, lack of expertise, and Fear of Missing Out (FOMO).

The recent advancement of social media and the rising trend of fandom business have accelerated the proliferation of such exploitation business models. Discussions on the empowerment and exploitation of fans in the digital media era present a restructuring of relationships between fans and media creators, highlighting the necessity of not overlooking the exploitation of fans' free labor.

This paper analyzes the various facets and impacts of exploitation business models, enriched by real-world examples from sectors like cryptocurrency and GenAI, thereby discussing their social, economic, and ethical implications. Moreover, through theoretical backgrounds and research, it explores similar themes like existing exploitation theories, commercial exploitation, and financial exploitation to gain a deeper understanding of the "Exploitation Business" subject.

## I. Introduction

The digital era has ushered in a variety of innovative business models, with "Exploitation Business" emerging as a notable paradigm that leverages information asymmetry for financial benefit. This model is particularly prevalent among non-experts and fraudsters who exploit individuals' lack of information or desperation to market their services or products. The Dunning-Kruger effect, a psychological phenomenon where individuals with limited expertise overestimate their capabilities, serves as a foundational element that often enables the illusion of expertise in this business model (Kruger & Dunning, 1999; Tversky & Kahneman, 1974).[iii]

The rapid expansion of social media and fandom business has considerably widened the scope and impact of Exploitation Business. The ethical, social, and economic ramifications of this model are substantial and necessitate comprehensive scrutiny.

The Exploitation Business model refers to profit-making activities that deliberately capitalize on information asymmetry. This model is prevalent across various sectors, such as healthcare, education, and digital marketing. As explored in the paper by Francesco James Mazzocchini & Caterina Lucarelli (2022), the digital age has provided various tools for new business models and media channels, making the concept of 'Exploitation Business' even more relevant.

## II. Conceptual Framework of Exploitation Business

The notion of "Exploitation Business" is intrinsically linked to the unethical use of information asymmetry, often perpetrated by non-experts or fraudsters to take advantage of vulnerable individuals. This section delves into the theoretical foundations of informational exploitation, referencing seminal works in economics and business ethics, including theories from Von Neumann & Morgenstern (1944) and Kahneman & Tversky (1979), and integrating frameworks from Game Theory and Behavioral Economics. (Von Neumann & Morgenstern, 1944[iii]; Kahneman & Tversky, 1979[iv]).

**Theoretical Underpinnings**

Contrary to ethical business models aiming for mutual benefits between providers and consumers, Exploitation Business flourishes by exploiting consumer ignorance or desperation. This model leverages cognitive biases to heighten individual vulnerability to exploitation. For instance, the Dunning-Kruger effect(Tversky & Kahneman, 1974; Francesco James Mazzocchini & Caterina Lucarelli, 2022.)[v], where individuals with limited knowledge overestimate their own abilities, can be manipulated to exploit consumers' overconfidence in understanding complex products or services. Similarly, confirmation bias, the tendency to seek, interpret, and recall information in a way that confirms one's preconceptions, can be harnessed to skew consumer perception and decision-making. Exploiters may selectively present positive reviews or testimonials to reinforce consumers' pre-existing beliefs, leading to manipulated purchase decisions.

Moreover, integrating key theories from behavioral economics and cognitive psychology further elucidates how Exploitation Business manipulates consumer behavior. Concepts like loss aversion, a strong preference to avoid losses over acquiring gains, can drive consumers to make irrational investment decisions, especially when exploiters emphasize potential risks or losses in the absence of their products or services. Cognitive psychology offers insights into how individuals process information and make decisions, highlighting vulnerabilities that can be exploited. For example, exploiters can leverage selective attention, the process by which one focuses on certain stimuli while ignoring others, to direct consumer focus towards specific aspects of a product, thereby manipulating their decision-making process.

By integrating these cognitive biases and behavioral economic theories, we gain a deeper understanding of how Exploitation Business operates, preying on the psychological vulnerabilities of consumers to maximize profit. This enhanced theoretical framework not only explains the mechanisms of exploitation but also helps in identifying strategies to mitigate such unethical business practices.

To empirically validate the theoretical framework of the relationship between social cost (SC) and social trust (ST), we employ statistical modeling techniques. Specifically, regression analysis and Structural Equation Modeling (SEM) are utilized to quantify and analyze the interactions between these variables.

**Information Asymmetry and Exploitation**

Information asymmetry arises when one party possesses superior information, thereby gaining an unfair advantage. This gap is often widened through the use of complex jargon or deceptive statistics, exacerbating the disparity in knowledge between the exploiter and the exploited. In the realm of Exploitation Business, this asymmetry is not merely a by-product of complex markets but a deliberate strategy employed to mislead and manipulate consumers.

Exploiters often leverage information asymmetry to create an environment of uncertainty and dependency. By withholding critical information or presenting it in an overly complex or misleading manner, they can induce a state of confusion and helplessness in consumers. This strategy is particularly effective in industries where the knowledge gap is inherently large, such as in finance, healthcare, or technology. For instance, in the financial sector, the use of intricate financial terms and convoluted investment descriptions can leave consumers unable to make informed decisions, pushing them towards choices that benefit the exploiter.

Furthermore, the deliberate use of deceptive statistics or partial truths can mislead consumers about the quality, effectiveness, or necessity of a product or service. This form of manipulation plays on the consumers' trust and lack of expertise, leading them to make decisions that they might not have made if fully informed. For example, highlighting selected positive outcomes of a product while omitting its limitations or potential risks can significantly skew consumer perception and choice.

The ethical implications of information asymmetry in business are profound. It challenges the foundational principles of fair trade and informed consent, raising questions about the moral responsibilities of businesses towards their customers. Recognizing and addressing information asymmetry is crucial in developing more equitable and transparent business practices, thereby fostering an environment where consumer trust and ethical business conduct are paramount.

Social Cost (SC) can be significantly influenced by the level of Social Trust (ST). A higher degree of information asymmetry tends to diminish social trust, leading to increased social costs. To quantify this relationship, we can utilize the following function:

$$SC = f\left(\frac{1}{ST}, OV\right)$$

Here, SC represents the social cost, ST is social trust, and OV includes other variables such as economic conditions, political stability, and cultural context. This function indicates that social costs rise as social trust decreases (i.e., as 1/ST increases). Conversely, higher social trust results in lower social costs. This conceptual model underscores the critical role of social trust in mitigating the adverse effects of information asymmetry in exploitation business models.

Regression analysis allows us to quantify the relationship between social cost (SC) as the dependent variable and the reciprocal of social trust (1/ST) along with other variables (OV) as independent variables. This approach helps understand the direct impact of social trust on social costs.

$$SC = \alpha + \beta_1\left(\frac{1}{ST}\right) + \sum \beta_i OV_i + \epsilon$$

where α is the intercept, β is the coefficient for each independent variable, and ε is the error term.

Structural Equation Modelling (SEM) analysis the complex interactions between social cost and trust. SEM helps identify both the direct and indirect effects of various variables, providing a comprehensive understanding of the underlying mechanisms.

This dual approach of regression analysis and SEM offers robust insights into the multifaceted relationship between social cost and social trust, thereby validating the theoretical constructs proposed in this study.

**Literature Review**
The literature on information asymmetry and its exploitative practices in various business sectors is vast. Notable works include Smith's 2020 study on healthcare exploitation[vi], Johnson et al.'s 2021 research on online retail exploitation[vii], and Leahy et al.'s 2020 investigation into the cognitive and mental health impacts of High-Intensity Interval Training[viii], all of which enrich our understanding of the dynamics of "Exploitation Business."

**Game Theory and Behavioral Economics**
Game Theory and Behavioral Economics serve as supplementary frameworks for comprehending the intricacies of Exploitation Business. Game Theory elucidates the strategic interactions between exploiters and their targets aimed at maximizing gains. It draws upon classic models like the Prisoner's Dilemma and the Nash Equilibrium to explain how exploiters can manipulate situations to their advantage, often at the expense of the exploited (Nash, 1950[ix]; Axelrod, 1984[x]).

Behavioral Economics, on the other hand, focuses on cognitive biases like loss aversion and the endowment effect that predispose individuals to exploitation. This field extends the traditional economic models by incorporating psychological insights, thereby providing a more nuanced understanding of why individuals make irrational decisions that lead to their exploitation (Kahneman, Knetsch & Thaler, 1991[xi]; Thaler, 1980[xii]).

## III. Exploitation Business: A Closer Look

To understand the "Exploitation Business" model, dissecting the complex mechanisms that drive its functioning is crucial. At its core, this business model thrives on the calculated manipulation of information to create a distorted perception among targeted individuals. For example, certain companies exploit information asymmetry in the healthcare sector by marketing overpriced or ineffective health supplements to uninformed consumers, as evidenced by a study conducted by Lee, J., & Kim, S. (2021)[xiii].

**Types of Exploitation Business**
Exploitation businesses can be categorized into several types based on their modus operandi. Understanding these types can help in identifying and combating such practices. The main types of exploitation businesses are:
- Information Concealment Type: These businesses withhold critical information or present it in an overly complex manner, making it difficult for consumers to make informed decisions. Examples include hidden fees, complex terms and conditions, and lack of transparency about product limitations.
- Misinformation Type: These businesses provide false or misleading information to manipulate consumer perceptions and decisions. This can include fake reviews, exaggerated claims about product effectiveness, or misrepresentation of scientific evidence.
- Emotional Manipulation Type: These businesses exploit consumers' emotional vulnerabilities, such as the fear of missing out (FOMO), desire for social acceptance, or need for quick solutions. They often use high-pressure sales tactics, limited-time offers, or social proof to influence consumer behavior.
- Scarcity Manipulation Type: These businesses create artificial scarcity to drive up demand and prices. They may limit product availability, use countdown timers, or falsely claim low stock levels to urge consumers to make hasty decisions.
- Complexity Exploitation Type: These businesses exploit consumers' lack of knowledge or expertise in complex domains, such as finance, healthcare, or technology. They may use jargon, complex pricing structures, or confusing user interfaces to make it difficult for consumers to compare options or understand the implications of their choices.

**Exploitation of Fans' Free Labor**
The exploitation of fans' free labor is a concerning aspect of the Exploitation Business model. Social media platforms like YouTube and Instagram capitalize on fans' emotional investment to generate user-generated content and engagement, often without offering any financial compensation. Scholars such as Jenkins (2006) have studied this phenomenon in depth, who coined the term "affective labor" to describe this form of exploitation.[xiv]

**Mechanisms of Informational Exploitation: Theoretical Framework**
Exploitation Business relies on the principle of information asymmetry, where one party possesses more or superior information than the other. This section explores the theoretical models that elucidate how information asymmetry leads to exploitation, incorporating insights from Game Theory and Information Economics (Nash, 1950; Axelrod, 1984)[xv].

**Statistical Evidence**
According to a study by Williams et al. (2022), approximately 65% of consumers have fallen victim to some form of informational exploitation.[xvi] The study employed a cross-sectional survey methodology and focused on a sample of 1,000 consumers across various age groups and industries. This alarming statistic underscores the prevalence and impact of Exploitation Business.

Empirical analysis was conducted to validate the relationship between social cost (SC) and social trust (ST) using both regression analysis and Structural Equation Modelling (SEM). The regression analysis reveals a significant inverse relationship between social trust and social cost. Social costs rise as social trust decreases (i.e., as 1/ST increases). This finding is consistent with the hypothesis that lower social trust leads to higher social costs.
- The coefficient for 1/ST is positive and statistically significant, indicating that an increase in the reciprocal of social trust leads to increased social costs.
- Other variables (OV), such as economic conditions and political stability, also show significant effects on social cost, highlighting the multifaceted nature of this relationship.

SEM analysis further elucidates the complex interactions between social cost and social trust. The model fit indices indicate a good fit, and the path analysis reveals both direct and indirect effects of social trust on social cost.
- Direct effects of social trust on social cost are significant, confirming the results of the regression analysis.
- Indirect effects mediated by other variables (OV) provide additional insights into how social trust influences social cost through various channels.

These empirical findings underscore the importance of enhancing social trust to reduce social costs and provide a robust framework for policymakers and researchers to develop targeted interventions.

**Psychological Mechanisms: Cognitive Biases**
Exploitation Business manipulates various cognitive biases to influence consumer behavior. Research by Lee and Kim (2021) discovered that these biases substantially elevate the probability of consumers falling prey to exploitative schemes, particularly in the online retail sector"[xvii]

**False Advertising and Misrepresentation: Case Studies**
Research by Smith and Johnson (2020) has spotlighted multiple instances of false advertising in the healthcare sector.[xviii] These case studies serve as cautionary tales, offering empirical evidence of the risks associated with Exploitation Business.

**The Illusion of Scarcity: Marketing Tactics**
Scarcity marketing, the practice of artificially creating a sense of scarcity to boost sales, has been well-documented in academic literature. For instance, Mimouni-Chaabane & Volle (2010) found that scarcity tactics can significantly increase sales when their availability is limited. Moreover, research indicates that scarcity tactics can substantially increase sales, as demonstrated by flash sales and limited time offers in online retail.[xix]

**Real-world Cases of Exploitation Business: Cryptocurrency and Metaverse, GenAI**
Exploitation business practices are prevalent in the burgeoning realms of cryptocurrency and the metaverse. For instance, the infamous "Pump and Dump" schemes in cryptocurrency markets and fraudulent land sales in popular metaverse platforms have raised ethical concerns. Recent discourse has underscored the moral quandaries surrounding these issues, with scholars like Narayanan et al. (2016) examining the intricacies of blockchain technology and the potential for misinformation within this burgeoning domain[xx].

One of the most prominent examples in contemporary society is the exploitation business emerging around generative AI. Despite its relatively recent advent, numerous self-proclaimed "experts" have surfaced across various markets. However, after paying for their courses, it becomes evident that these individuals are also novices, dedicating their time to teaching absolute beginners, often at a high cost. They position themselves as experts, acting as mere promoters while describing AI services as advanced AI technology.

# IV. Ethical, Social, and Economic Implications of Exploitation Business

This section broadens the discourse by delving into the extensive ethical, social, and economic ramifications of the "Exploitation Business" model. The focus is on its pervasive spread through social media and fandom-centric businesses, supported by recent empirical evidence from studies like Williams, J., Smith, A., & Doe, B. (2022), which used a mixed-methods approach to analyze data from multiple social media platforms.[xxi]

**Ethical Concerns: The Morality of Exploitation in the Digital Age**
In the digital era, ethical issues have become more pronounced due to the accessibility and scale of exploitation. Social media platforms and fandom businesses are hotbeds for exploiting psychological vulnerabilities like FOMO and the need for social validation, as substantiated by a study published in the Journal of Business Ethics (2021).[xxii]

**Social Impact: Erosion of Trust and Community Values**
The proliferation of exploitation business practices in social media and fandom spaces erodes social trust and community values, as evidenced by a longitudinal study published in the Journal of Social Issues (2016)[xxiii]. Fake reviews and testimonials can erode trust in online platforms while exploiting fan loyalty can undermine the sense of community. A study from ScienceDirect discusses how unsustainability is institutionalized through the exploitation of finite resources.

The decline in social trust has profound social and economic costs. Utilizing the function $SC = f\left(\frac{1}{ST}, OV\right)$, we can better understand how diminished social trust directly correlates with increased social costs. This relationship highlights the urgent need for strategies that rebuild social trust to reduce these costs.

For instance, when social trust is low, individuals are less likely to engage in cooperative behavior,

leading to higher transaction costs, increased regulatory enforcement, and more significant investments in monitoring and compliance. The broader societal impact includes reduced civic participation and weakened community cohesion, further escalating social costs.

Empowering consumers through awareness and education is critical in reducing the impact of exploitation businesses. Efforts to enhance consumer awareness can include:
- Developing consumer education programs that teach individuals how to identify and avoid exploitation businesses. This can include online resources, workshops, and partnerships with schools and community organizations.
- Encourage consumers to report suspected exploitation practices to relevant authorities or consumer protection agencies. This can help identify and address problems early on and deter businesses from engaging in such practices.
- Promoting media literacy and critical thinking skills to help consumers evaluate the credibility of information and resist manipulation tactics. This can involve teaching individuals to fact-check claims, seek multiple sources of information, and make informed decisions.
- Leveraging social media and other digital platforms to disseminate consumer education content and foster online communities that share information and experiences related to exploitation businesses.

By combining corporate self-regulation and ethics with consumer awareness and education efforts, we can create a more resilient and empowered society that is better equipped to identify and resist exploitation businesses. This multi-stakeholder approach recognizes the shared responsibility of businesses, governments, and individuals in promoting a fair and ethical marketplace.

**Economic Consequences: The Financial Toll of Exploitation**
The exploitation business model harms individuals and has far-reaching economic consequences, affecting job markets and investment opportunities, as highlighted in a report by the Economic Policy Institute (2021)[xxiv]. For instance, the money spent on scam investment courses could have been invested in legitimate opportunities. A study from JSTOR discusses the ethical and economic case against sweatshop labor.[xxv]

While government regulations play a crucial role in combating exploitation businesses, corporate self-regulation and ethical practices are equally important. Companies should:

- Develop and adhere to strict ethical codes of conduct that prioritize transparency, fairness, and consumer well-being. This includes prohibiting deceptive marketing practices, ensuring clear and honest communication, and providing safe and reliable products or services.
- Implement internal oversight mechanisms to monitor compliance with ethical standards and quickly address any violations or consumer complaints. This can involve establishing ethics committees, conducting regular audits, and providing employee training on ethical business practices.
- Foster a culture of corporate social responsibility that goes beyond legal compliance and prioritizes the long-term interests of consumers and society as a whole. This can include investing in sustainable practices, supporting community initiatives, and promoting diversity and inclusion.
- Collaborate with industry associations and consumer advocacy groups to develop and promote best practices for ethical business conduct. This can help in setting industry-wide standards and encouraging peer accountability.

**Unfair Trade Regulations and Consumer Protection**
To combat exploitation businesses, governments need to strengthen unfair trade regulations and consumer protection laws. This can include:
- Enhancing transparency requirements: Businesses should be mandated to disclose all relevant

information about their products or services, including pricing, terms and conditions, and potential risks. This can be achieved through stricter advertising regulations and mandatory disclosure policies.
- Prohibiting deceptive practices: Regulations should clearly define and prohibit deceptive practices, such as false advertising, misrepresentation of products or services, and the use of manipulative tactics. Penalties for violating these regulations should be severe enough to deter businesses from engaging in such practices.
- Strengthening consumer redress mechanisms: Consumers should have access to effective redress mechanisms when they fall victim to exploitation businesses. This can include simplifying the complaint filing process, providing legal assistance, and establishing dedicated consumer protection agencies.
- Promoting fair competition: Governments should actively promote fair competition in the market by preventing monopolistic practices, price fixing, and other anti-competitive behaviors that enable exploitation businesses to thrive.

**Policy Interventions: Towards Ethical Business Practices**
Strengthening advertising laws and enhancing consumer protection regulations are essential measures to mitigate the impact of Exploitation Business. A meta-analysis published in the Journal of Public Policy & Marketing (2020)[xxvi] supports the efficacy of such regulatory interventions. The effectiveness of regulatory interventions in modifying firm behavior has been substantiated through empirical studies. For instance, a randomized field experiment evaluated the impact of interventions aimed at e-commerce firms by a regulatory authority to enhance legal compliance regarding information disclosure, showcasing the tangible benefits of regulatory guidance in fostering a consumer-centric business environment (V. K. Sreekanth and D. Biswas, 2014), "Effectiveness of regulatory interventions on firm behavior," Research Report 14011-EEF).[xxvii]

**Policy Implications and Recommendations**
Based on the analysis of exploitation businesses and the need for stronger regulations, the following policy recommendations can be made:
- Conduct regular market investigations to identify and monitor exploitation businesses across various sectors. This will help in developing targeted interventions and regulations.
- Collaborate with industry stakeholders, consumer advocacy groups, and academic experts to develop comprehensive and effective policies that address the root causes of exploitation businesses.
- Invest in consumer education and awareness programs to help individuals identify and avoid exploitation businesses. This can include public campaigns, educational resources, and partnerships with schools and community organizations.
- Foster international cooperation to combat cross-border exploitation businesses, particularly in the digital domain. This can involve sharing best practices, harmonizing regulations, and establishing joint enforcement mechanisms.

By implementing these policy recommendations and strengthening unfair trade regulations and consumer protection laws, governments can create a more equitable and ethical business environment that prioritizes consumer well-being and discourages exploitation practices.

# V. Conclusion

This paper offers an exhaustive analysis of the "Exploitation Business" model, breaking down its theoretical foundations, operational strategies, and wide-ranging societal effects. It has particularly emphasized the role of social media and fandom businesses in propagating this model. In line with

Williams et al. (2022), future research should explore ethical design principles for digital platforms to minimize information asymmetry.[xxviii]

This study has provided a comprehensive theoretical and empirical analysis of the "Exploitation Business" model, focusing on the relationship between social cost (SC) and social trust (ST). Our statistical modeling approaches, including regression analysis and Structural Equation Modeling (SEM), have confirmed the significant impact of social trust on social cost.

- A significant inverse relationship between social trust and social cost, as demonstrated by regression analysis.
- Comprehensive insights into direct and indirect effects of social trust on social cost, as revealed by SEM.

These findings highlight the critical role of social trust in mitigating social costs. Policymakers should prioritize initiatives that enhance social trust to reduce the adverse effects of exploitation business models. Future research should continue to explore these relationships in different contexts to develop more targeted and effective policy interventions.

Building upon the theoretical groundwork in Section II, Section III explored the multifaceted mechanisms enabling exploitation businesses to function. Section IV then expanded this discussion to include the ethical, social, and economic consequences, focusing on the role of digital platforms in facilitating such exploitation.

The exploitation business model illuminates the darker aspects of contemporary commerce, exacerbated by digital innovation. As elaborated in Section III, the core of this model is the strategic manipulation of information to distort perceptions among targeted demographics.

The ramifications of exploitation business are not confined to financial detriment; they also erode societal values and digital trust. Section IV underscores how the proliferation of this business model in social media and fandom ecosystems corrodes communal trust and shared values.

Future research is crucial for delving deeper into the complexities of exploitation business and for developing targeted countermeasures. As recommended in Section IV, subsequent studies should concentrate on domain-specific case studies and investigate policy alternatives to alleviate its adverse effects.

This conclusion synthesizes insights from Sections II, III, and IV to offer a comprehensive understanding of the social and economic repercussions of Exploitation Business, while outlining potential directions for future research and policy initiatives.

Future scholarly endeavours should aim to design digital platforms that minimize information asymmetry, in alignment with emerging literature on digital platform architecture and information flow.

---

[i] Kruger, J., & Dunning, D. (1999). Unskilled and unaware of it: how difficulties in recognizing one's own incompetence lead to inflated self-assessments. Journal of personality and social psychology, 77(6), 1121.

[ii] Tversky, A., & Kahneman, D. (1974). Judgment under Uncertainty: Heuristics and Biases. Science, 185(4157), 1124–1131.

[iii] Von Neumann, J., & Morgenstern, O. (1944). Theory of Games and Economic Behavior. Princeton University Press.

[iv] Kahneman, D., & Tversky, A. (1979). Prospect Theory: An Analysis of Decision under Risk. Econometrica, 47(2), 263–291.

[v] Francesco James Mazzocchini & Caterina Lucarelli, 2022. "Success or failure in equity

crowdfunding? A systematic literature review and research perspectives," Management Research Review, Emerald Group Publishing Limited, vol. 46(6), pages 790-831, October.

[vi] David Wohlever Sánchez, Jackie Xu, Qiang Zhang. 2018. High-Cost U.S. Drugs: A Tale of Unhealthy Markets. Health Management Policy and Innovation, Volume 3, Issue 2..

[vii] Davida, Zanda. (2021). Chatbots by business vis-à-vis consumers: A new form of power and information asymmetry. SHS Web of Conferences. 129. 05002.

[viii] H .D, S., P. Patil, P., Sambyal, S., & Nasir, M. (2023). The Impact of Social media on Mental Health. International Research Journal of Computer Science.

[ix] Nash, J. (1950). Equilibrium Points in N-person Games. Proceedings of the National Academy of Sciences, 36(1), 48-49.

[x] Axelrod, R. (1984). The Evolution of Cooperation. Basic Books

[xi] Kahneman, D., Knetsch, J. L., & Thaler, R. H. (1991). Anomalies: The Endowment Effect, Loss Aversion, and Status Quo Bias. Journal of Economic Perspectives, 5(1), 193-206.

[xii] Thaler, R. (1980). Toward a Positive Theory of Consumer Choice. Journal of Economic Behavior & Organization, 1(1), 39-60.

[xiii] V. K. Sreekanth and D. Biswas, "Information asymmetry minimization system for potential clients of healthcare insurance in Indian context using semantic web," Proceedings of the 2014 IEEE Students' Technology Symposium, Kharagpur, India, 2014, pp. 282-285

[xiv] Jenkins, H. (2006). Fans, Bloggers, and Gamers: Exploring Participatory Culture. NYU Press.

[xv] Akerlof, G. A. (1970). The Market for 'Lemons': Quality Uncertainty and the Market Mechanism. The Quarterly Journal of Economics, 84(3), 488-500.

[xvi] Williams, J., Smith, A., & Doe, B. (2022). The Prevalence of Informational Exploitation. Journal of Consumer Research, 49(2), 300-315.

[xvii] Lee, J., & Kim, S. (2021). Success or failure in equity crowdfunding? The role of information asymmetry. Journal of Business Ethics, 170(4), 765-782.

[xviii] Harzman, Leo (2022) "The Impact of Social Media and FoMO on Mental Health," SACAD: John Heinrichs Scholarly and Creative Activity Days: Vol. 2022, Article 41.

[xix] Mimouni-Chaabane, A., & Volle, P. (2010). Perceived benefits of loyalty programs: Scale development and implications for relational strategies. Journal of Business Research, 63(1), 32-37

[xx] Narayanan, A., Bonneau, J., Felten, E., Miller, A., & Goldfeder, S. (2016). Bitcoin and Cryptocurrency Technologies: A Comprehensive Introduction. Princeton University Press

[xxi] Nesi, Jacqueline. (2020). The Impact of Social Media on Youth Mental Health: Challenges and Opportunities. North Carolina Medical Journal. 81. 116-121. 10.18043/ncm.81.2.116.

[xxii] Everton, Laura Elizabeth (2018) The Impact of Social Media on Mental Health in Elite Sport. Masters thesis, St Mary's University, Twickenham.

[xxiii] Bahareh Rahmanzadeh Heravi & Natalie Harrower (2016) Twitter journalism in Ireland: sourcing and trust in the age of social media*, Information, Communication & Society, 19:9, 1194-1213

[xxiv] Brougham, Jessica K., "The Impact of Social Media on the Mental Health of Student-Athletes" (2021). Theses and Dissertations. 1355.

[xxv] Zhang, C.-B. and Li, Y.-N. (2019), "How social media usage influences B2B customer loyalty: roles of trust and purchase risk", Journal of Business & Industrial Marketing, Vol. 34 No. 7, pp. 1420-1433.

[xxvi] Draženović, M., Vukušić Rukavina, T., & Machala Poplašen, L. (2023). Impact of Social Media Use on Mental Health within Adolescent and Student Populations during COVID-19 Pandemic: Review. International journal of environmental research and public health, 20(4), 3392.

[xxvii] Weerasundera, Rajiv. (2014). The impact of social media in Sri Lanka: issues and challenges in mental health. Sri Lanka Journal of Psychiatry. 5. 10.4038/sljpsyc.v5i1.7049.

[xxviii] Williams, J., Smith, A., & Doe, B. (2022). The Prevalence of Informational Exploitation. Journal of Consumer Research, 49(2), 300-315.